\documentclass[]{interact}

\usepackage{xcolor}
\usepackage{multirow}
\usepackage{hyperref}

\usepackage{epstopdf}% To incorporate .eps illustrations using PDFLaTeX, etc.
\usepackage[caption=false]{subfig}% Support for small, `sub' figures and tables

\usepackage[numbers,sort&compress]{natbib}% Citation support using natbib.sty
\bibpunct[, ]{[}{]}{,}{n}{,}{,}% Citation support using natbib.sty
\theoremstyle{plain}% Theorem-like structures provided by amsthm.sty

\theoremstyle{definition}

\theoremstyle{remark}

\begin{document}

\articletype{General Application}% Specify the article type or omit as appropriate

\title{Spatial similarity index for scouting in football}

\author{
\name{V. Gómez-Rubio\textsuperscript{a}, J. Lagos\textsuperscript{b} and F. Palmí-Perales\textsuperscript{c}\thanks{CONTACT F. Palmí-Perales. Email: francisco.palmi@uv.es}}
\affil{\textsuperscript{a}Universidad de Castilla-La Mancha (Albacete, Spain); \textsuperscript{b}Scoutanalyst S.L.; \textsuperscript{c}Universitat de València (València, Spain)}
}

\maketitle

\begin{abstract}
Finding players with similar profiles is an important problem in sports such as football. Scouting for new players requires a wealth of information about the available players so that similar profiles to that of a target player can be identified. However, information about the position of the players in the field is seldom used. For this reason, a novel approach based on spatial data analysis is introduced to produce a spatial similarity index that can help to identify similar players. The use of this new spatial similarity index is illustrated to identify similar players using spatial data from the Spanish competition ``La Liga'', season 2019-2020.
\end{abstract}


\section{Introduction}

Scouting in sports refers to looking for players with a particular profile \cite{Johnstonetal:2018}. It can be conducted to fill a particular position within the team or to replace another player that might be innjured or leaving the team. Hence, scouting plays an important role in sports management as well.

Typically, scouting can be performed by outlining the required skills, quantifyng them and looking for players with similar values of these skills. Nowadays,
different sports databases and websites will provide accurate statistics about different players that can be used for scouting \cite{LawlorPalmer:2023}.

Spatial data about the players is currently available on a regular basis and it is often used to produce heatmaps to show the players position during a game. This is often closely related to the role of the player within the team. However, this spatial information is seldom used to identify similar profiles of players according to their positions in the field.

Football data is usually obtained by either eventing or tracking
\cite{Papalardoetal:2019,Papalardoetaldata:2019}. Eventing refers to recording
information about any relevant event (hence, the name) during the game. For
example, when a foul is commited, which is a relevant event during a game,
variables recorded will include the player that produces the foul, player that
receives the foul, position in the field, etc.  Alternatively, tracking refers
to recording information using different devices (includong TV broadcast).  For
example, the position of a player may be recorded using a small GPS so that
other variables such as distance covered during a game, etc. In general, data collected in eventing will require more human intervention than the one obtained with tracking.

Recent reviews on scouting have surveyed the current state of research in this field \citep{ScoReview2, ScoReview1}, discussing key findings and developments. In addition, other authors \citep{ScoFuture} propose several promising yet unexplored research directions. The primary focus of this study, however, is on quantitative analysis rather than qualitative evaluations, which have also been addressed in prior research \citep{qualit1}.

In terms of quantitative analysis, a number of studies employ machine learning techniques to derive insights from extensive datasets. For instance, \citep{ScoIA} utilizes different methods to develop an application that compiles statistics for individual players and leverages machine learning and rule-based models to examine player relationships and statistical interdependencies. Similarly, \citep{ScoChap} employs data mining and predictive modeling to evaluate player performance using a variety of machine learning strategies available through the Rminer package. 

Machine learning is also applied in additional studies such as \citep{ScoMLPos}, where a random forest algorithm is used to simulate and assess a new player's fit within a team. This approach essentially creates a variable representing the player’s “fit” within the team, offering a means of quantifying the player's “value” through a single composite indicator. This approach aligns with other quantitative studies in scouting that focus on developing composite indicators for ranking players, such as \citep{ScoIndica1, ScoIndica2}. 

Other quantitative methodologies in scouting include the use of multiple regression models to analyze relationships between specific scouting variables \citep{ScoMultiRegress}. However, more complex models that analyze players' performance through position-tracking data are relatively uncommon. For example, \citep{ScoSpa} models pass effectiveness by analyzing player coordinates on the field. The study develops a composite indicator for passes and fits a Ridge regression model to explain variability in this indicator, observing associations between pass location, speed, and angle, though the model’s overall fit was limited. 

As seen in the aforementioned study, spatial modeling has been explored by various researchers in football. Some studies focus on modeling player and ball positions using motion models \citep{Motions1, Motions2}. Other studies have employed alternative methods to capture spatial and temporal aspects of field dynamics without employing a fully spatio-temporal model. For instance, passing trends have been modeled using motion models \citep{Motions3} or basic linear models such as ANOVA \citep{PassANOVA}. Furthermore, position-tracking data has been used to analyze player or team positions through descriptive statistics and the development of various metrics \citep{track1, track2}. 

While there are multiple approaches to incorporating spatial coordinates into the modeling process, many studies do not employ formal spatial models or spatial techniques. Instead, they tend to model coordinates or centroids using alternative methods. However, a few studies propose models within a spatial framework. For example, \citep{NMSTPP} presents a model for the timing, location, and types of in-game football events based on a Transformer-Based Neural Marked Spatio-Temporal Point Process (NTPP), which combines point process modeling with machine learning techniques.

Alternatively, \citep{narayanan2023flexible} proposes a Bayesian framework for inference, offering a comprehensive model of flexible marked point processes suitable for analyzing sequences of in-game events in football. This model allows for the simulation of in-game events and estimation of event probabilities, enabling calculations of expected goal value at specific moments within a game. Despite these contributions, spatial methodologies in football analysis remain scarce, with few studies offering formal spatial modeling approaches.

Given the lack of methods to consider spatial data in football for scouting, a
spatial similarity index is proposed to address this issue by exploiting the
spatial information available about the players. In particular, the spatial
information will be represented using a lattice of the field so that it is
divided into small squares or rectangles for which a measure of the occupation
or activity of the player is available (e.g., the average time spent during a
game). Then, these values will be compared using appropriate spatial
cross-correlation statistics such as Lee's statistic
\cite{Lee:2004,Mateoetal:2016}, from which a spatial similarity index is
derived. Although the computation of the spatial similarity index will be
illustrated with an example from football \cite{Beggs:2024}, it is suitable for
similar sports.

This paper is organized as follows. Section~\ref{sec:spatial} provides
an introduction to spatial data analysis. Methods to assess spatial similarity are discussed in Section~\ref{sec:spsim}. Next, an example is developed
in Section~\ref{sec:example}. Finally, a discussion and some comcluding remarks are provided in Section~\ref{sec:discussion}.

\section{Spatial data analysis}
\label{sec:spatial}

Spatial data analysis refers to a particular area of statistics that deals with data for which a location is available. Several authors \cite{Haining:2003,Bivandetal:2013} provide a detailed summary of the different types of data and the corresponding statistical methodology to analyse them.

In this particular case, the spatial domain will be the field and it will be
assumed that it has been divided into smaller regular areas (usually squares or
rectangles), so that a number of relevant variables have been recorded for each
area.  Typically, this will be the average number of minutes spent in each area
or, alternatively, the proportion of time spent in each area.

\begin{figure}[h]
\centering
\includegraphics[height=7cm]{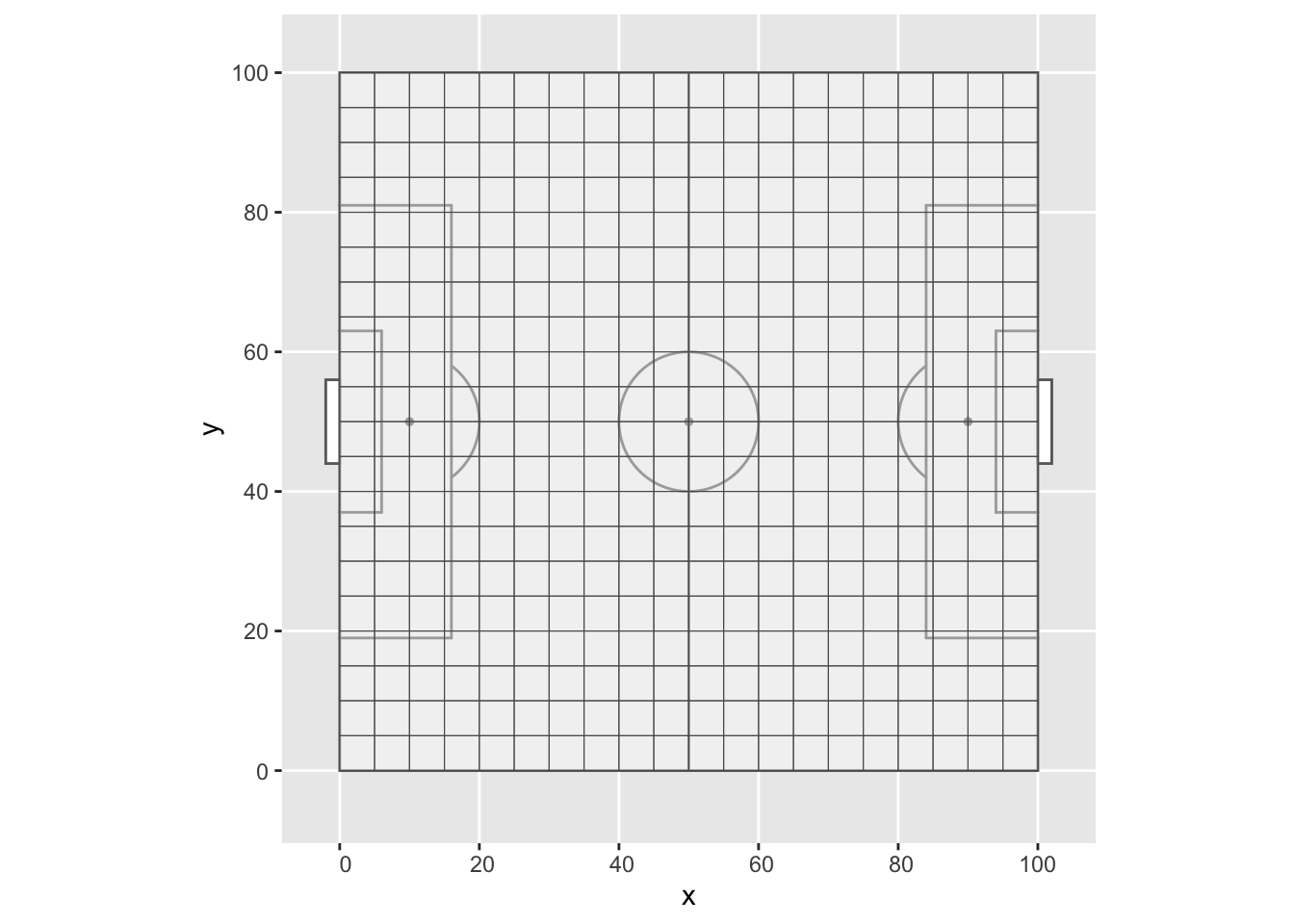}\includegraphics[height=7cm]{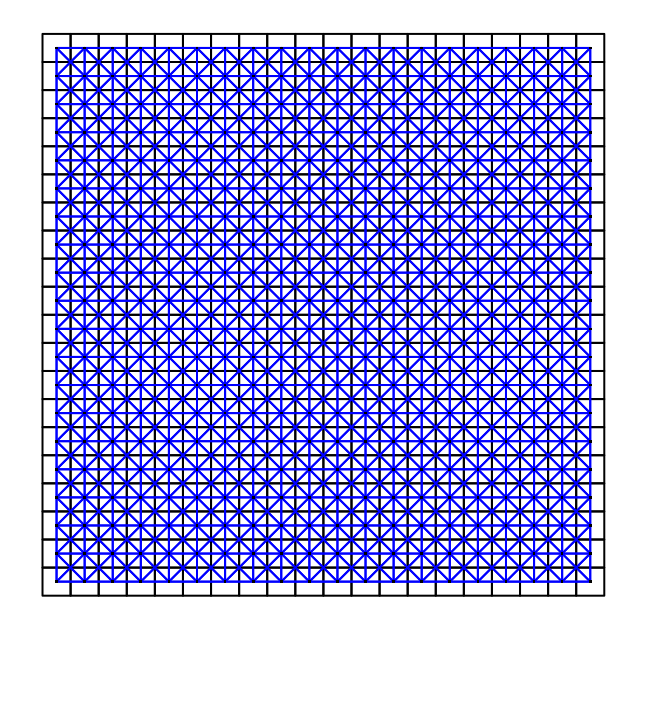}
\caption{Spatial representation of a football field using small areas (left) and representation of the adjacency (blue lines) between the different small areas (right).}
\label{fig:spatial}
\end{figure}

Figure~\ref{fig:spatial} illustrates how a football field can be split into smaller regular areas (top plot), using the R package \citep{ggsoccer}. Note that each area is represented by a polygon and there are a number of variables that could be recorded. For example, the number of yellow or red cards issued by an action in each area, average number of players in each area and many others. For the current analysis, the focus will be in the time spent by each player in each area and developing a suitable methodology to assess their similarities. Furthermore, Figure~\ref{fig:spatial} (right plot)
includes a representation of the adjacency between the different small areas.

Figure~\ref{fig:toyexample} shows heatmaps for 
5 different players.  From a statistical point of view, this representation
can be regarded as a summary of the available data.  From a scouting
perpsective, given a target player with a particular spatial profile the aim is
to find the player with a very similar spatial profile.

When dealing with spatial data, the variables represented will often show some
spatial pattern, pointing to the fact that values from neighbouring regions are
not independent. This means that values from areas that share a boundary (i.e.,
neighbours) will have similar values.  This spatial autocorrelation can be
tested in different ways, as exaplined inn Section~\ref{subsec:spauto}.

Regarding neighbourhood, it can be represented as a binary matrix with as many rows and columns as the number of areas so that entries with a value of 1 mean that areas $i$ and $j$ are neighbours and an entry of 0 means that they are not.
In addition to neighbourhood, spatial weights can be defined between any pair of areas to measure the the strength of their relationship. However, most often spatial weights will be taken as 1 and 0, in a similar way as neighbourhood. However, other alternatives are avilable \cite{Bivandetal:2013}. The matrix of spatial weights will be represented by $W$.

\subsection{Spatial autocorrelation}
\label{subsec:spauto}

A convenient way to test for spatial autocorrelation is Moran's I statistic.
It is defined as

$$
I = \frac{n}{\sum_{i=1}^n \sum_{j=1}^n w_{ij}}
  \frac{\sum_{i=1}^n \sum_{j=1}^n w_{ij}(x_i-\bar{x})(x_j-\bar{x})}{\sum_{i=1}^n (x_i - \bar{x})^2}
$$
\noindent
Here, $n$ refers to the number of areas, $w_{ij}$ is the spatial weight
between areas $i$ and $j$, $x_i$ is the value of the variable of interest in
area $i$ and $\bar{x}$ the mean of $\{x_i\}_{i=1}^n$.

Moran's I will be positive when either areas with values over the mean $\bar{x}$ tend to have neighbours with also values over the mean or when areas with values below the mean tend to have neighbours with values also below the mean. When
the oposite happens, the value of Moran's I will be negative. When there is no spatial structure in the data, the value of Moran's I will be close to zero.

Hence, positive values of Moran's I indicate positive spatial association. A statistical test can be develop to assess whether the value of Moran's I
statistic is significantly high \cite{Bivandetal:2013}. 

It is important to note that the presence of spatial autocorrelation will
depend on the nature of the phenomenon under study. In our particular case, it
is reasonable to find spatial autocorrelation because players move smoothly thourought the field and tend to occupy different parts of the field depending on their role in the team.

\subsection{Spatial cross-correlation}
\label{subsec:Lee}

Moran's I statistic has been introduced in order to outline the important
elements to consider when assessing spatial autocorrelation. However, Moran's I
canot deal with more than one variable. In order to compare the spatial
distribution of two variables, Lee's statistic will be used.

Given a set of $n$ areas with associates weights $\{w_{ij}\}_{i,j=1}^n$, where
variables $\{x_i\}_{i=1}^n$ and $\{y_i\}_{i=1}^n$ have been measured, Lee's
statistic is defined as follows:

$$
L= \frac{n}{\sum_{i=1}^{n}(\sum_{j=1}^{n}w_{ij})^2} 
  \frac{\sum_{i=1}^{n}\left(\sum_{j=1}^{n}w_{ij}(x_i-\bar{x})\right) \left(\sum_{j=1}^{n}w_{ij}(y_j-\bar{y})\right)}{\sqrt{\sum_{i=1}^{n}(x_i - \bar{x})^2} \sqrt{\sum_{i=1}^{n}(y_i - \bar{y})^2}}
$$

Note that Lee's statitic will have positive values when both variables tend to
have values above or below their respective mean values, similarly as it
happened with Moran's I statistic. Lee's statistic will provide negative values
when one variable tend to have values above the mean and their neighbours tend
to have values below the mean (or viceversa).

Furthermore, a statistical test using the expectation and variance of the
statistics under spatial independence. Note that the alternative hypothesis of interest is that of positive spatial correlation as this implies that two players tend to occupy similar areas in the field. A negative spatial association (which will yuield a negative value of Lee's statistic) is not of interest because this implies that two players will tend to occupy different positions in the field.

\section{Spatial similarity}
\label{sec:spsim}

In order to provide a measure of spatial similarity, Lee's statistic
\cite{Lee:2004} will be used to assess spatial association between the spatial
distribution of each pair of players. Note that this statistic can be computed
on a  number of variables, e.g., average time spent or proportion of time in each area, and
that this willl lead to values of the test statistic in different scales. Note
also that different players may be more prone to play and, hence, their values
may be higher, leading also to a changes of scale when computing Lee's statistic. For this reason, it is advisable to compute Lee's statistic using the proportion of time spent in each square of the field. 

As a metric to measure spatial similarity between two players, the standardized
Lee's statistic (i.e., $L$ minus its expectation divided by the standard
deviation) or the associated p-value of a one-sided test (see
Section~\ref{subsec:Lee}) could be used.  However, to have a bounded measure of
association that in independent from the scale, the aforementioned p-value will
be used. In this way, values close to zero will indicate a strong spatial
association while values close to 1 will mean that there is either no spatial
association or this is negative. In any of these cases, the two players will be
similar in terms of their spatial position in the field.

To sum up, the associated p-value can be regarded as a \textit{pseudo-distance} in the
space of players. It is worth noting that this is not a distance in the
strict mathematical sense but a pseudo-measure of dissimilarity, i.e, the higher the value
the less similar the players are. If a similarity measure is needed, then one
minus the p-value can be taken.

Once the similarity measure has been obtained, players could be grouped together by means of different clustering algorithms so that a hierarchical clustering can be conducted \cite{Everitt:1974}. This hierarchical clustering will be useful in order to group players together. 

\section{Example: spatial distribution of soccer players}
\label{sec:example}

Two examples will be developed in this Section to illustrate the methods
developed in previous sections. First of all, an example with only 5 players (from Spanish competition ``La Liga'', season 2019-2020) is included in Section~\ref{subsec:toyex} to provide some insight on how the proposed methodology works. Next, an example with ten of thousands of players
is discussed in Section~\ref{subsec:fullex} to illustrate a real application of the proposed methodology. Note that due to confidentiality reasons only the datasets for the toy examples are available.

Data for the examples below have been adquired from data provider Wyscout
(\url{https://www.hudl.com/en_gb/products/wyscout}). This is a subset of a
major dataset that corresponds to players from different leagues worldwide,
season 2019-2020. Altogether, the original dataset comprises 14914 players from 1389 teams worldwide. However, for the purpose of this paper, only the 500 players from
20 teams in the Spanish football competetion ``La Liga'', season 2019-2020,
are considered.

Data comes as a set of coordinates that represent the centroids of some
squares in the field after transforming the game field into a square in $[0,100]\times[0,100]$. Attached to each coordinate is a value that represents
the activity of the player in that area of the field (e.g., average time spent in that square).

In order to match the data from the different players to the same regular grid
(see Figure~\ref{fig:spatial}, left plot) to compute the similarity measure, a
weighted density smoothing will be conducted using the centroids and the
associated value as weights so that a raster image is created for all players.
This will ensure that all players' data are represented using the same raster
image so that the regular lattice and its adjacency structure can be computed.

The analysis has been conducted on a Linux cluster running Ubutu with 64 cores Intel(R) Xeon(R) CPU E5-2683 v4 @ 2.10GHz and 157 Gb of RAM. The main R packages used in the analysis have been spatstat \cite{spatstat} and  sp \cite{sp_pkgs:2022} for spatial data handling and vizualisation and spdep \cite{SDS:2023} to compute Lee's statistic and the associated spatial test and p-value. R package \texttt{factoextra} \cite{factoextra} has been used to visualize the distances between the players.

\subsection{Toy example}
\label{subsec:toyex}

For this first example, data about 5 different players from the Spanish competition ``La Liga'' (season 2019-2020) will be used. These players are Unai Simón (goalkeeper, Atlético de Bilbao), Éver Banega (midfilder, Sevilla F. C.), Toni Kroos (midfilder, Real Madrid), Dani Parejo (midfilder, Valencia F. C.), and Lionel Messi (forward, F. C. Barcelona). These selection of players covers different roles and distributions on the field.

\begin{figure}[h!]
\centering
\includegraphics[scale=0.3]{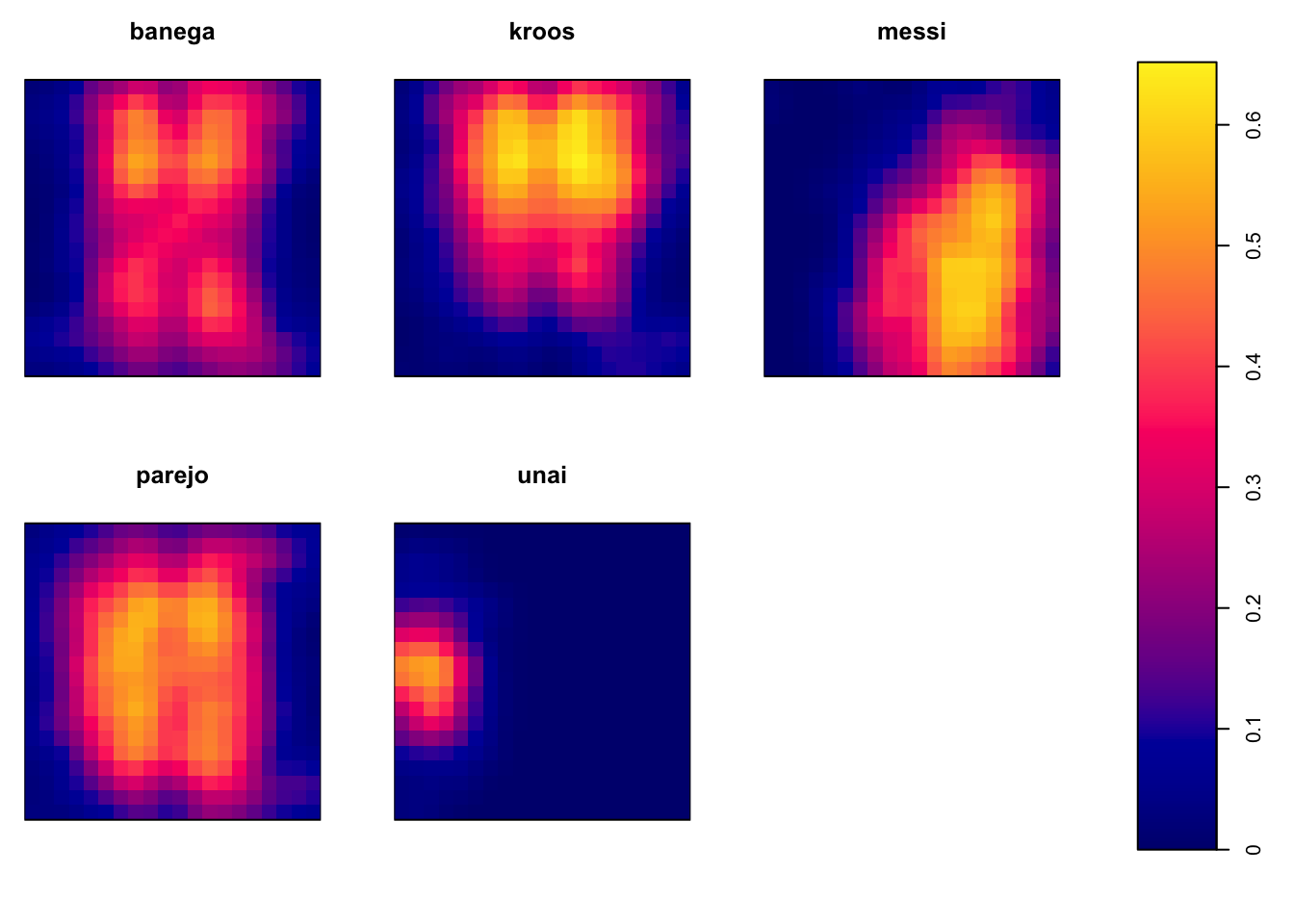}
\caption{Spatial distribution of 5 different players from ``La Liga'', season 2019-2020.}
\label{fig:toyexample}
\end{figure}

Figure~\ref{fig:toyexample} shows the different spatial distributions of the players. As it can be seen, there are clear differences between some of them while there are also clear similarities among the midfilders.

In order to provide some insight on the proposed methodology to identify similar players, Table~\ref{tab:toyexample} shows the values of Lee's statistic when comparing each pair of players (diagonal and lower triangle) as well as the associated p-value of the one-sided test (upper triangle).

\begin{table}[h!]
\centering
\begin{tabular}{|l|l|rrrrr|}
\hline
& & \multicolumn{5}{|c|}{\textbf{Player}}\\
\hline
& & banega & kroos & messi & parejo & unai\\
\hline
\multirow{2}{*}{\textbf{Player}} & banega &  49.92 &  0 & 0 & 0 & 1\\
& kroos &   42.78 & 53.98 & 0 &  0 & 1\\
& messi &   15.42 &   3.77 & 56.49 &  0 & 1\\
& parejo &  42.05 &  36.01 & 15.87 & 49.70 &  1 \\
& unai &   -22.72 & -15.72 & -27.52 & -7.81 & 52.13 \\
\hline
\end{tabular}
\caption{Values of Lee's statistic (diagonal and lower triangle) and associated p-value from one-sided test (upper triangle) for 5 different players from ``La Liga'', season 2019-2020.}
\label{tab:toyexample}
\end{table} 

Note how the values of Lee's statistic when a player is compared to himself are quite large, but note also that these are different from one player to another. These values lead to very small p-values which, in practice, can be taken as zero.

Regarding the comparisons between two different players, Unai has a negative Lee's statistic with all the other players given his disctintive spatial distribution in the field. In particular, all values of Lee's statistic are negative when compared to other players as field occupancy is eseentiallt different to all the other players compared. This means a negative spatial association and to a ver small p-value. 

When comparing all the other players, Lee's statistic is always positive and p-values are very small. However, by looking at the values of the statistic, it is possible to observe that all three midfilders (i.e., Banega, Kroos and Parejo) have larger values among them than whe compared with Messi (who is a forward). 

\begin{figure}[h!]
\centering
\includegraphics[scale=0.3]{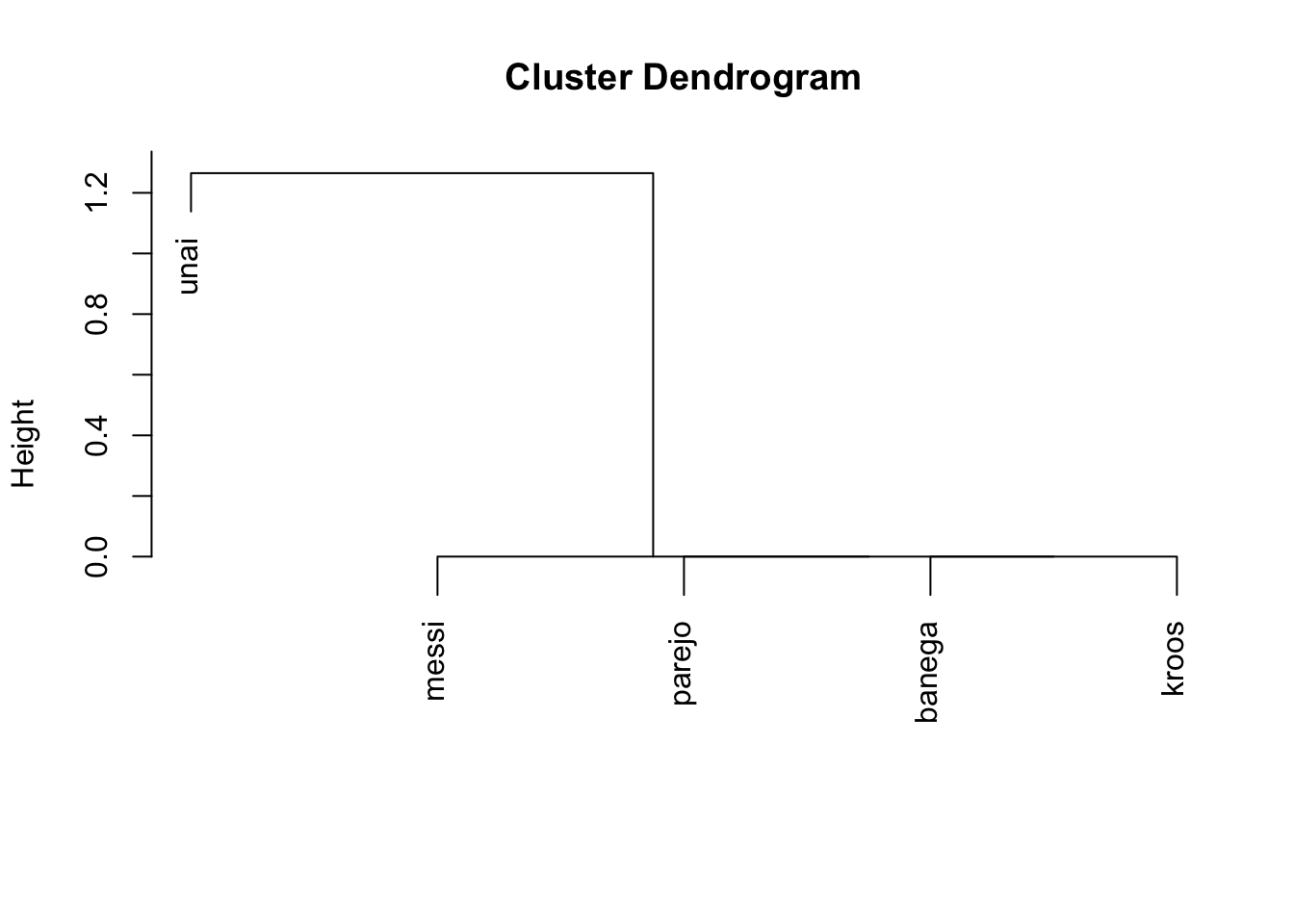}
\caption{Dendrogram of a hierarchical clustering for 5 different players from ``La Liga'', season 2019-2020.}
\label{fig:toycluster}
\end{figure}

Next, a hierarchical clustering will be conducted in order to determie the
different groups of players in the dataset. The algorithm starts with each
player being a different cluster and then clusters are aggregated in turn so
that the two most similar clusters are combined in each step until all players
are together in one cluster. The distance between two clusters is the largest
distance between any player in the first cluster and any player in the second
one. This is immplemented in the R function \texttt{hclust()} when the
\texttt{method} argument is set to \texttt{"complete"} (the default).
Note that considering other distances between the cluster may lead to slightly different clusterings.

Figure~\ref{fig:toycluster} shows the dendrogram from the hierarchical
clustering using the pseudo-distance computed using the p-value from Lee's
statistic.  As it can be seen, the hierarchical clustering seems to identify
two different groups according to the different roles of the players in the
example.

Note also that the pseudo-distances are very close to all for all players but
Unai, and that when he is compared to all other players the pseudo-distance is very close to 1 in all cases. This makes the dendrogram difficult to interpret as all pseudo-distances are, in practice, either zero or one. In fact, they are not exactly zero or one but the tiny differences cannot be appreciated in Table~\ref{tab:toyexample}.

However, by inspecting how the clustering is done, more insight on how the
pseudo-distance are used can be gained. First, Banega and Kroos are put
together in the first iteration.  This is not surpirsing as Banega and Kroos
have the highest value of Lee's statistic as well as the the smallest p-value.
In sucessive iterations, Parejo, Messi and Unai are added (in this order) to
the previous cluster. Again, Parejo is added first because he is the most
similar player to Banega and Kroos as he is also a midfielder. He is followed
then by Messi (a forward) and, finally, by Unai (a goalkeeper).

Although this is a small example, it has illustrated how the proposed
methodology to compare and cluster players according to their spatial
distributio in the field works. Next, a more realistic example comprising a
lager number of players and teams is developed.

\subsection{Analysis of ``La Liga'', season 2019-2020}
\label{subsec:fullex}

In order to illustrate the use of the spatial similarity index, it will be
applied to the players in the Spanish competition ``La Liga'', season 2019-2020,
which is the highest level football league in Spain. This
requires comparing and computig the index for each pair of players. However, this is something that can be easily done in parallel and the computations of the index are simple.

Furthermore, not all players need to be compared at the same time and they ca be grouped according to their role. For example, goalkeepers do not need to be compared with any other players. Also, different variables can be considered prior to computing the spatial similarity index. However, to best illustrate the performance of the proposed methodology, all players will be compared to each other.

\begin{figure}[h!]
\centering
\includegraphics[scale=0.15]{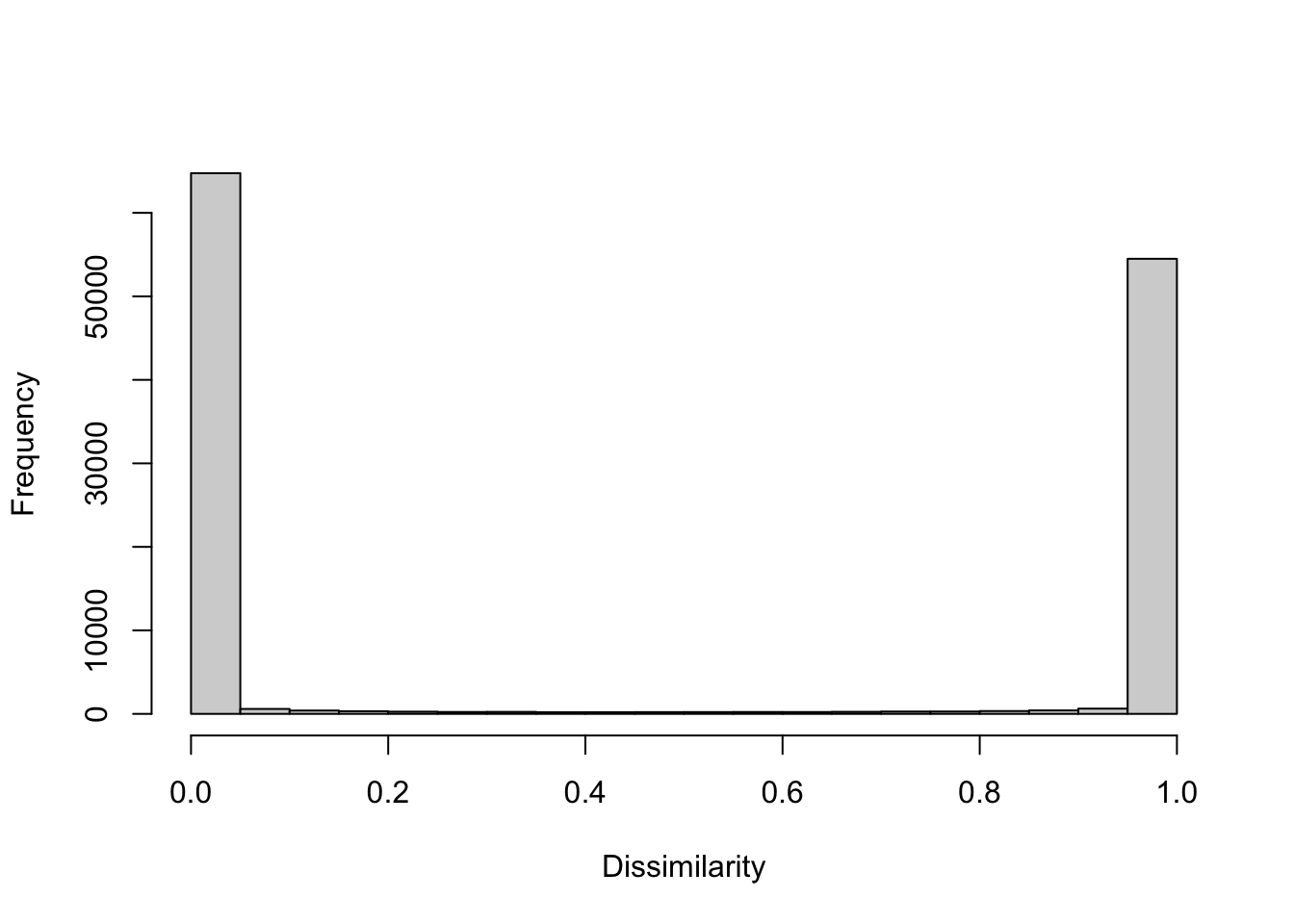}\,\,\includegraphics[scale=0.15]{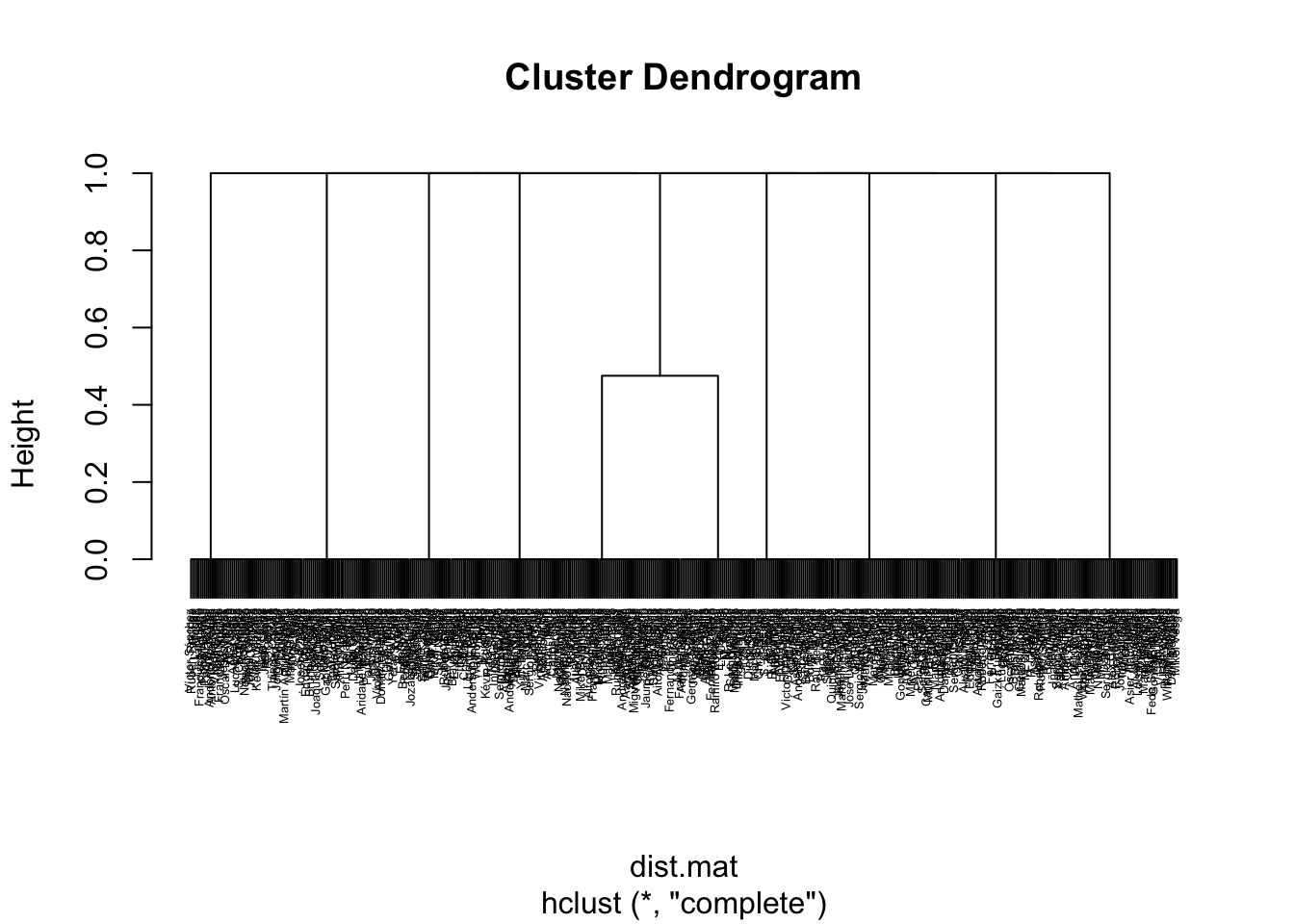}
\caption{Distribution of spatial similarity index values (left) and resulting dendrogram of a hierarchical clusteirng (right) for all player considered from `La Liga'', season 2019-2020.}
\label{fig:SSI}
\end{figure}

Hence, the similarity index has been computed using the methodology described in Section~\ref{sec:spsim} and a hierarchical clustering computed. 
The left plot in Figure~\ref{fig:SSI} summarizes the values of the
pseudo-distances obtained when comparing all the players. As it can be seen,
most of the values are very close to either 0 or 1, with just a few in between.
This points to the fact that Lee's statistic is quite strict about detecting
similar patterns.

As stated above, a hierarchical clustering has been conducted on the dataset
and it can be seen in the right plot in Figure~\ref{fig:SSI}. The dendrogram
seems to define 10 main groups, with different number of players in each of
them.  Furthermore, setting a cut-off point of 0.001 seems reasonable in order
to break the players down into groups. As a result, 10 groups of players are
obtained, with an average number of  players of 50 and a range between 21 to
75.

\begin{figure}[h!]
\centering
\includegraphics[scale=0.15]{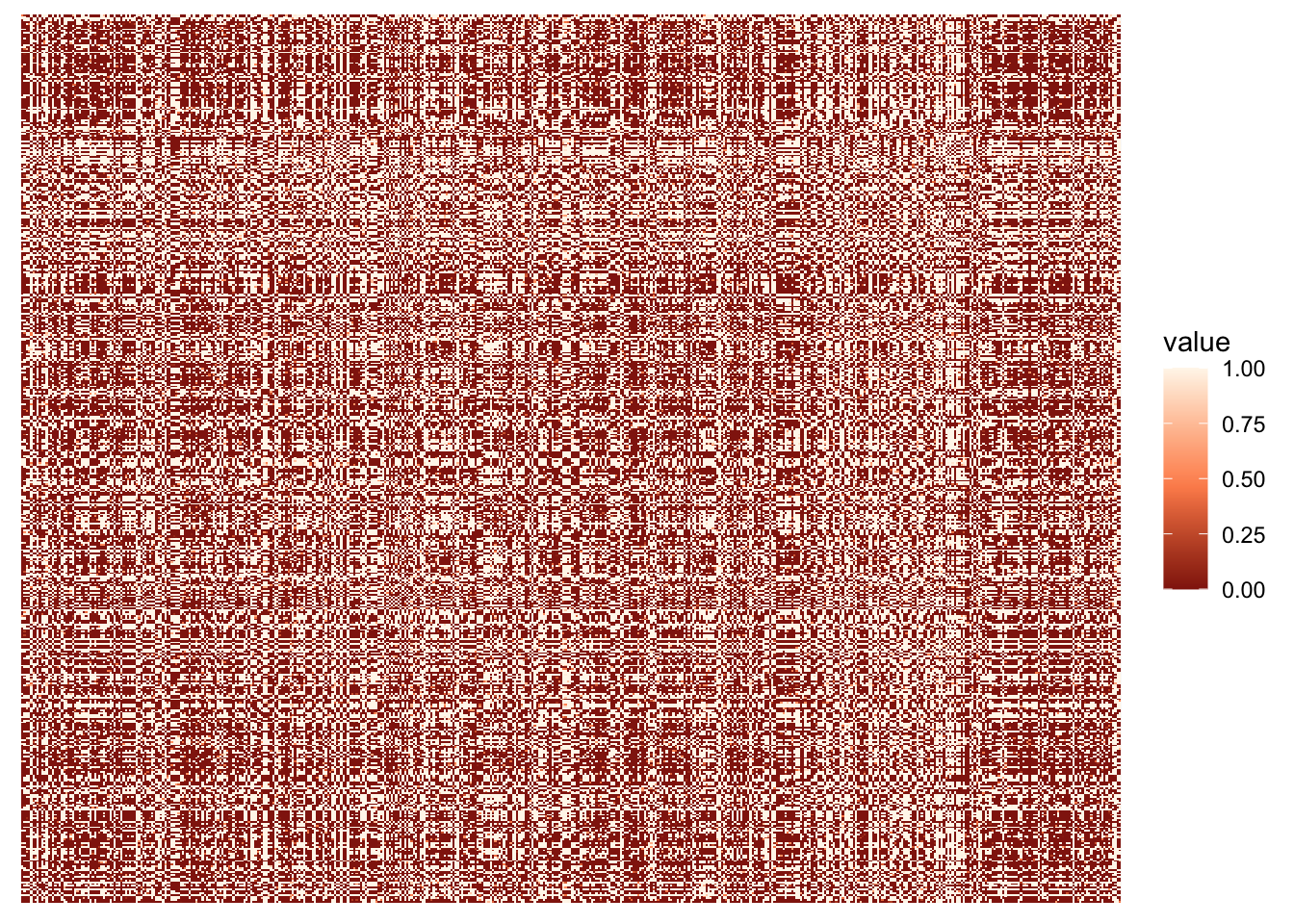}\,\,\includegraphics[scale=0.15]{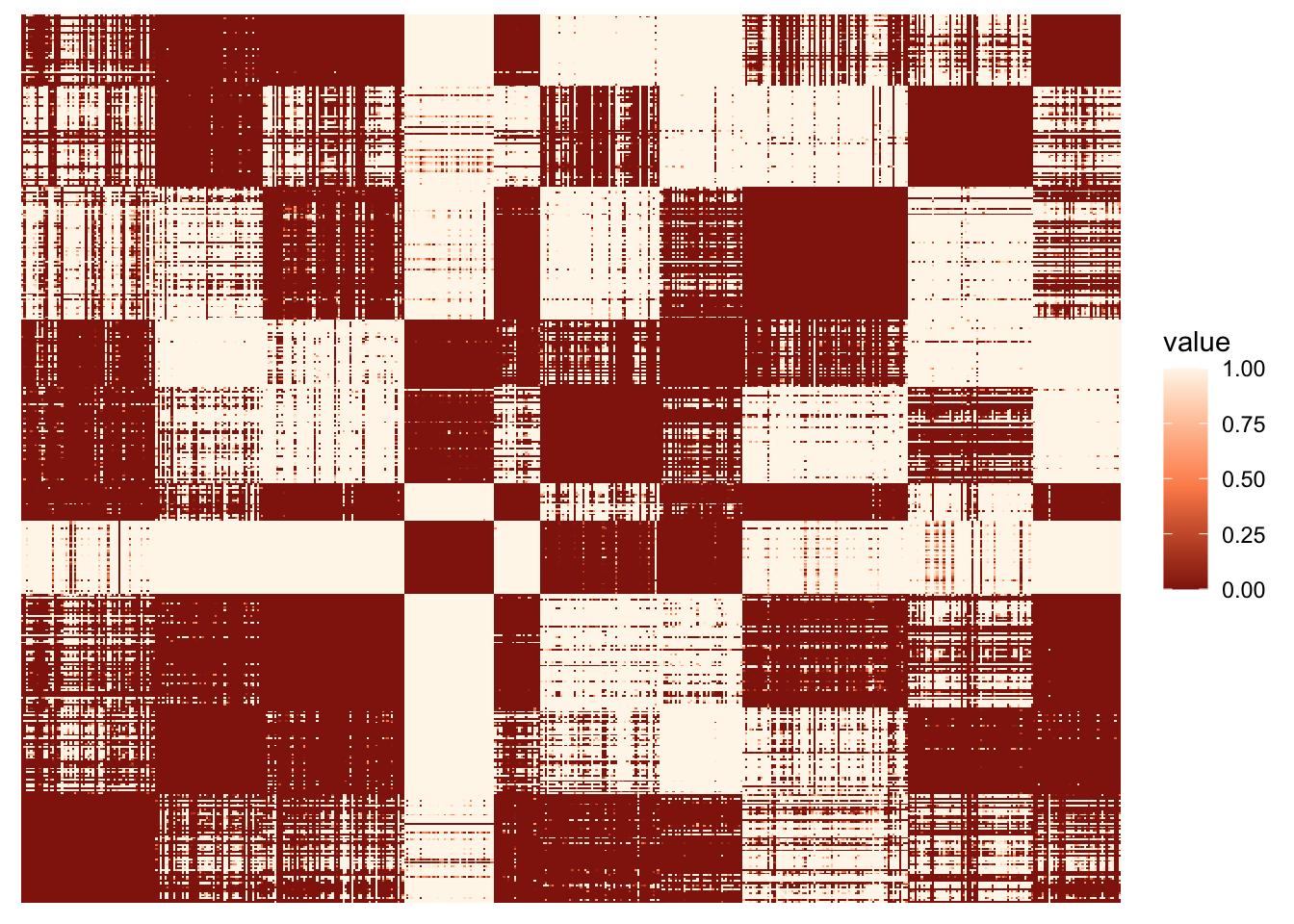}
\caption{Matrix of similarities between players in the original order of the data (left) and after reordering according to the clusters found (right).}
\label{fig:dist}
\end{figure}

Similarly, the left plot in Figure~\ref{fig:dist} shows the distance between each pair of players in the original data, which shows no pattern at all as players are originally ordered by team. However, the right plot in Figure~\ref{fig:dist} shows the same distances after the players have been sorted according to their cluster number. In this plot, 10 blocks can be easily identified in the minor diagonal (from down-left to top-right). The plots have been produced with R
package \texttt{factoextra} \citep{factoextra}.

\begin{figure}[h!]
\centering
\includegraphics[scale=0.8]{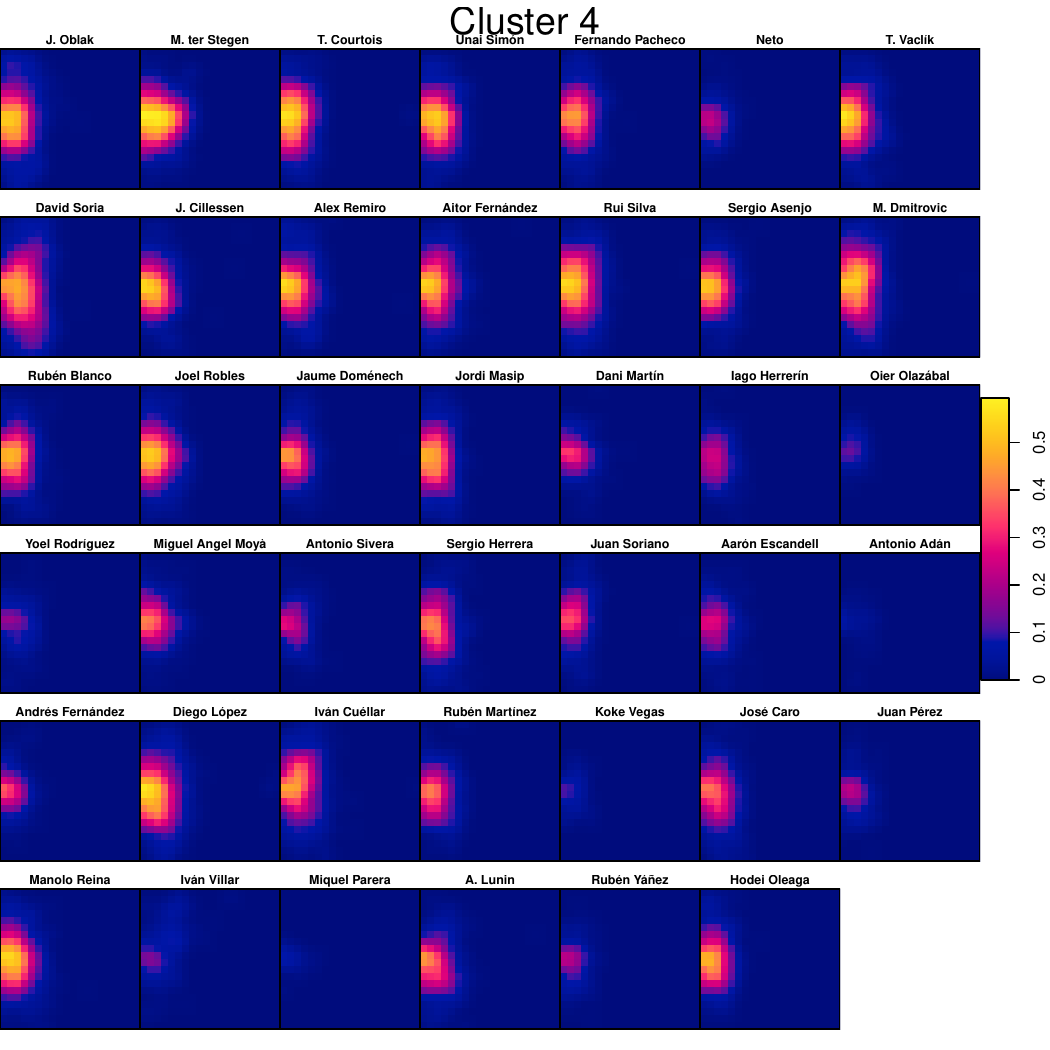}
\caption{Heatmaps of players in cluster 4, which clearly identifies the goalkeepers.}
\label{fig:cluster4}
\end{figure}

Regarding the clusters, they seem to identify clearly the different types
of players in the data. The heatmaps of the players in the different clusters have been included as Supplementary Material. However, in Figure~\ref{fig:cluster4} the heatmaps of players in cluster 4 has been included. As it can be seen,
this cluster clearly identifes the goalkeepers in the dataset.

For these reasons, it is possible to conclude that the similarity index is
a suitable metric to group players according to their position
in the field during the game.

\section{Discussion and concluding remarks}
\label{sec:discussion}

As it could not be otherwise, the role of a player in the team in intimetly linked to its position in the field. Hence, by exploring the spatial distribution it is possible to obtain a sensible metric for comparing different players. Players with similar spatial distributions will, most likely, have similar roles within their respective teams. Hence, by comparing these spatial distributions to find similarities and dissimilarities it will be possible to match similar players. This is particularly important to replace injured players or to look for similar players in the market.

The methodology described in this paper can effectively discriminate among a larg number of players according to their spatial distribution in the firld during the game. Note that this spatial similarity index can be used after players have been filtered according to other characteristics. Hence, by combining this spatial simialry index with other attributes, a more effective scouting can be achieved.

Note that the spatial variable measured can be the (average) time spent in each area of the field but it can also be taylored in different ways to match specific needs. For example, specific spatial profiles can be produced for each player when the teams is attacking and defending, so that the target player should now match these profiles rather than a single profile.

Similarly, when tracking data is used, the spatial profiles can
be used to inspect the position of the players that do not take part in the
main action with the ball. This can lead to off-ball movements that may be
important during the game. For example, a midfilder may occupy parts of the
field near the area where a potential pass can be received so that it takes
defenders away from where the main action is taking place.

Although the example develop in this paper describes an application to
football, the spatial similarity index and the methodology presented in this
paper could be used for other similar team sports. Furthermore, the spatial
similarity iindex could be combined with other existing data to better identify
similar profiles. 

%\section*{Acknowledgement(s)}
%
%\textbf{ESTO SEGURAMENTE LO DEBAMOS QUITAR.}
%
%Data has been provided by XXXXX, which have granted permission to share the datasets for the toy example.

\section*{Disclosure statement}

The authors declare no conflict of interest.

\section*{Funding}

This work has been supported by grant PID2022-136455NB-I00, funded by MCIN/AEI/10.13039/501100011033/FEDER and the European Regional Development Fund, grant CIAICO/2022/165, funded by Dirección General de Ciencia e Investigación (Generalitat Valenciana, Spain), and grant SBPLY/21/180501/000241, funded by Consejer\'ia de Educaci\'on, Cultura y Deportes (Junta de Comunidades de Castilla-La Mancha, Spain) and FEDER. 

\section*{Code availability}

Code and datasets to replicate the results in this paper are available at
\url{https://github.com/becarioprecario/spatialindex}. Note that raw data are not provided due to being commercial data but the heatmaps used to compute Lee's statistic are provided.

\bibliographystyle{tfs}
\bibliography{football}

\begin{thebibliography}{10}
\providecommand{\MR}{\relax\unskip\space MR }
\providecommand{\url}[1]{\normalfont{#1}}
\providecommand{\urlprefix}{Available at }

\bibitem{spatstat}
A. Baddeley, E. Rubak, and R. Turner, \emph{Spatial Point Patterns: Methodology
  and Applications with {R}}, Chapman and Hall/CRC Press, London, 2015,
  \urlprefix\url{https://www.routledge.com/Spatial-Point-Patterns-Methodology-and-Applications-with-R/Baddeley-Rubak-Turner/p/book/9781482210200/}.

\bibitem{Beggs:2024}
C. Beggs, \emph{Soccer Analytics. An Introduction Using {R}}, Chapman \&
  Hall/CRC Press, 2024.

\bibitem{Bivandetal:2013}
R.S. Bivand, E. Pebesma, and V. Gómez-Rubio, \emph{Applied Spatial Data
  Analysis with R}, 2nd ed., Springer, New York, 2013.

\bibitem{Everitt:1974}
B. Everitt, \emph{Cluster Analysis}, Heinemann Educational Books Ltd., London,
  1974.

\bibitem{ScoMLPos}
S. Ghar, S. Patil, and V. Arunachalam, \emph{Data Driven football scouting
  assistance with simulated player performance extrapolation}, in \emph{2021
  20th IEEE International Conference on Machine Learning and Applications
  (ICMLA)}. IEEE, 2021, pp. 1160--1167.

\bibitem{track1}
F.R. Goes, M.S. Brink, M.T. Elferink-Gemser, M. Kempe, and K.A. Lemmink,
  \emph{The tactics of successful attacks in professional association football:
  large-scale spatiotemporal analysis of dynamic subgroups using position
  tracking data}, Journal of Sports Sciences 39 (2021), pp. 523--532.

\bibitem{Motions1}
J. Gudmundsson and T. Wolle, \emph{Football analysis using spatio-temporal
  tools}, in \emph{Proceedings of the 20th International Conference on Advances
  in Geographic Information Systems}. 2012, pp. 566--569.

\bibitem{Haining:2003}
R. Haining, \emph{{Spatial Data Analysis: Theory and Practice}}, Cambridge
  University Press, 2003.

\bibitem{Johnstonetal:2018}
K. Johnston, N. Wattie, J. Schorer, and J. Baker, \emph{Talent identification
  in sport: A systematic review}, Sports medicine 48 (2018), pp. 97--109.

\bibitem{factoextra}
A. Kassambara and F. Mundt, \emph{factoextra: Extract and Visualize the Results
  of Multivariate Data Analyses} (2020).
  \urlprefix\url{https://CRAN.R-project.org/package=factoextra}, R package
  version 1.0.7.

\bibitem{ScoSpa}
M. Kempe, F.R. Goes, and K.A. Lemmink, \emph{Smart data scouting in
  professional soccer: Evaluating passing performance based on position
  tracking data}, in \emph{2018 IEEE 14th International Conference on e-Science
  (e-Science)}. IEEE, 2018, pp. 409--410.

\bibitem{LawlorPalmer:2023}
C. Lawlor and C. Palmer, \emph{‘being’ in the world of football scouting -
  an exercise in storied and performed data}, Journal of Qualitative Research
  in Sports Studies 17 (2023), pp. 107--126.

\bibitem{ScoFuture}
C. Lawlor, J. Rookwood, and C.M. Wright, \emph{Player scouting and recruitment
  in {English} men’s professional football: opportunities for research},
  Journal of Qualitative Research in Sports Studies 15 (2021), pp. 57--76.

\bibitem{Lee:2004}
S.I. Lee, \emph{A generalized significance testing method for global measures
  of spatial association: An extension of the mantel test}, Environment and
  Planning A: Economy and Space 36 (2004), pp. 1687--1703.
  \urlprefix\url{https://doi.org/10.1068/a34143}.

\bibitem{track2}
D. Link, S. Lang, and P. Seidenschwarz, \emph{Real time quantification of
  dangerousity in football using spatiotemporal tracking data}, PloS one 11
  (2016), p. e0168768.

\bibitem{Mateoetal:2016}
R. Mateo, O. Broennimann, S. Normand, B. Petitpierre, M. Araujo, J.C. Svenning,
  A. Baselga, F. Fernandez-Gonzalez, V. Gomez-Rubio, J. Munoz, G. Suarez, M.
  Luoto, A. Guisan, and A. Vanderpoorten, \emph{The mossy north: an inverse
  latitudinal diversity gradient in european bryophytes}, Scientific Reports 6
  (2016).

\bibitem{narayanan2023flexible}
S. Narayanan, I. Kosmidis, and P. Dellaportas, \emph{Flexible marked
  spatio-temporal point processes with applications to event sequences from
  association football}, Journal of the Royal Statistical Society Series C:
  Applied Statistics 72 (2023), pp. 1095--1126.

\bibitem{Motions2}
T. Narizuka, Y. Yamazaki, and K. Takizawa, \emph{Space evaluation in football
  games via field weighting based on tracking data}, Scientific reports 11
  (2021), p. 5509.

\bibitem{Papalardoetal:2019}
L. Pappalardo, P. Cintia, A. Rossi, E. Massucco, P. Ferragina, D. Pedreschi,
  and F. Giannotti, \emph{A public data set of spatio-temporal match events in
  soccer competitions}, Scientific Data 6 (2019).

\bibitem{Papalardoetaldata:2019}
L. Pappalardo and E. Massucco, \emph{Soccer match event dataset (version 2)}
  (2019).

\bibitem{ScoIA}
J. Pavitt, D. Braines, and R. Tomsett, \emph{Cognitive analysis in sports:
  Supporting match analysis and scouting through artificial intelligence},
  Applied AI letters 2 (2021), p. e21.

\bibitem{SDS:2023}
E. Pebesma and R.S. Bivand, \emph{Spatial Data Science With Applications in
  {R}}, Chapman \& Hall, 2023, \urlprefix\url{https://r-spatial.org/book/}.

\bibitem{ScoIndica1}
R. Poli, L. Ravenel, and R. Besson, \emph{Technical analysis of performances
  and player scouting}, CIES Football Observatory Monthly Report 60 (2020).

\bibitem{ScoIndica2}
F. Pretto, \emph{Development of a football analytics web application for player
  scouting}, Master's thesis, Escuela de Negocios,  2022.

\bibitem{qualit1}
E. Radicchi and M. Mozzachiodi, \emph{Social talent scouting: A new opportunity
  for the identification of football players?}, Physical Culture and Sport.
  Studies and Research 70 (2016), pp. 28--43.

\bibitem{sp_pkgs:2022}
 {Roger Bivand}, \emph{R packages for analyzing spatial data: A comparative
  case study with areal data}, Geographical Analysis 54 (2022), pp. 488--518.

\bibitem{ScoMultiRegress}
R. Soncin, J. Pennone, J.P. Pinho, M.C. Diniz, J.G. Claudino, A.C. Amadio, J.C.
  Serr{\~a}o, and B. Mez{\^e}ncio, \emph{Football scout analysis models (based
  in the 2013/2014 champions league)}, Revista Brasileira de Educa{\c{c}}{\~a}o
  F{\'\i}sica e Esporte 31 (2017), pp. 33--39.

\bibitem{Motions3}
B. Spencer, M. Hawkey, and S. Robertson, \emph{Using contextual player movement
  and spatial control to analyse player passing trends in football},
  Bar{\c{c}}a Sport. Analytics summit  (2019), pp. 1--12.

\bibitem{ggsoccer}
B. Torvaney, \emph{ggsoccer: Plot Soccer Event Data} (2024).
  \urlprefix\url{https://CRAN.R-project.org/package=ggsoccer}, R package
  version 0.2.0.

\bibitem{PassANOVA}
B. Travassos, R. Monteiro, D. Coutinho, F. Yousefian, and B. Goncalves,
  \emph{How spatial constraints afford successful and unsuccessful penetrative
  passes in elite association football}, Science and Medicine in Football 7
  (2023), pp. 157--164.

\bibitem{ScoChap}
T. Vilela, F. Portela, and M.F. Santos, \emph{Towards a pervasive intelligent
  system on football scouting-a data mining study case}, in \emph{Trends and
  Advances in Information Systems and Technologies: Volume 3 6}, Springer,
  2018, pp. 341--351.

\bibitem{NMSTPP}
C.C.K. Yeung, T. Sit, and K. Fujii, \emph{Transformer-based neural marked
  spatio temporal point process model for football match events analysis}
  (2023). \urlprefix\url{https://arxiv.org/abs/2302.09276}.

\bibitem{ScoReview1}
M. Yunus and R.S. Aditya, \emph{Talent scouting and standardizing fitness data
  in football club: systematic review}, Retos 62 (2024), pp. 1382--1389.

\bibitem{ScoReview2}
A. {\v{Z}}ivkovi{\'c}, \emph{et~al.}, \emph{The impact of information
  technologies on the scouting process in sports games}, in \emph{Sinteza
  2020-International Scientific Conference on Information Technology and Data
  Related Research}. Singidunum University, 2020, pp. 240--245.

\end{thebibliography}

\appendix

\section*{Supplementary Materials of paper ``Spatial similarity index for scouting in football''}

\section*{Clusters}

The paper describes how to cluster players according to a spatial index.
Altogether, players have been clustered into 10 different clusters.
A list of the different positions in the filed can be found in the
\href{https://en.wikipedia.org/wiki/Association_football_positions}{Wikipedia}
(for example).

After inspection of the clusters, they can be roughly described as follows:

\begin{itemize}

\item Cluster 1: Central midfilder.

\item Cluster 2: Right attacking midfilder.

\item Cluster 3: Center attacking midfilder.

\item Cluster 4: Goalkeeper.

\item Cluster 5: Left attacking midfilder.

\item Cluster 6: Right-back defender.

\item Cluster 7: Left-back defender.

\item Cluster 8: Left winger.

\item Cluster 9: Right winger.

\item Cluster 10: Forward.

\end{itemize}

\noindent
The heatmaps of the players abalyzed in the paper (Spanish competition ``La Liga'', season 2019-2020) is provided in the following pages.

\includegraphics[scale=0.9]{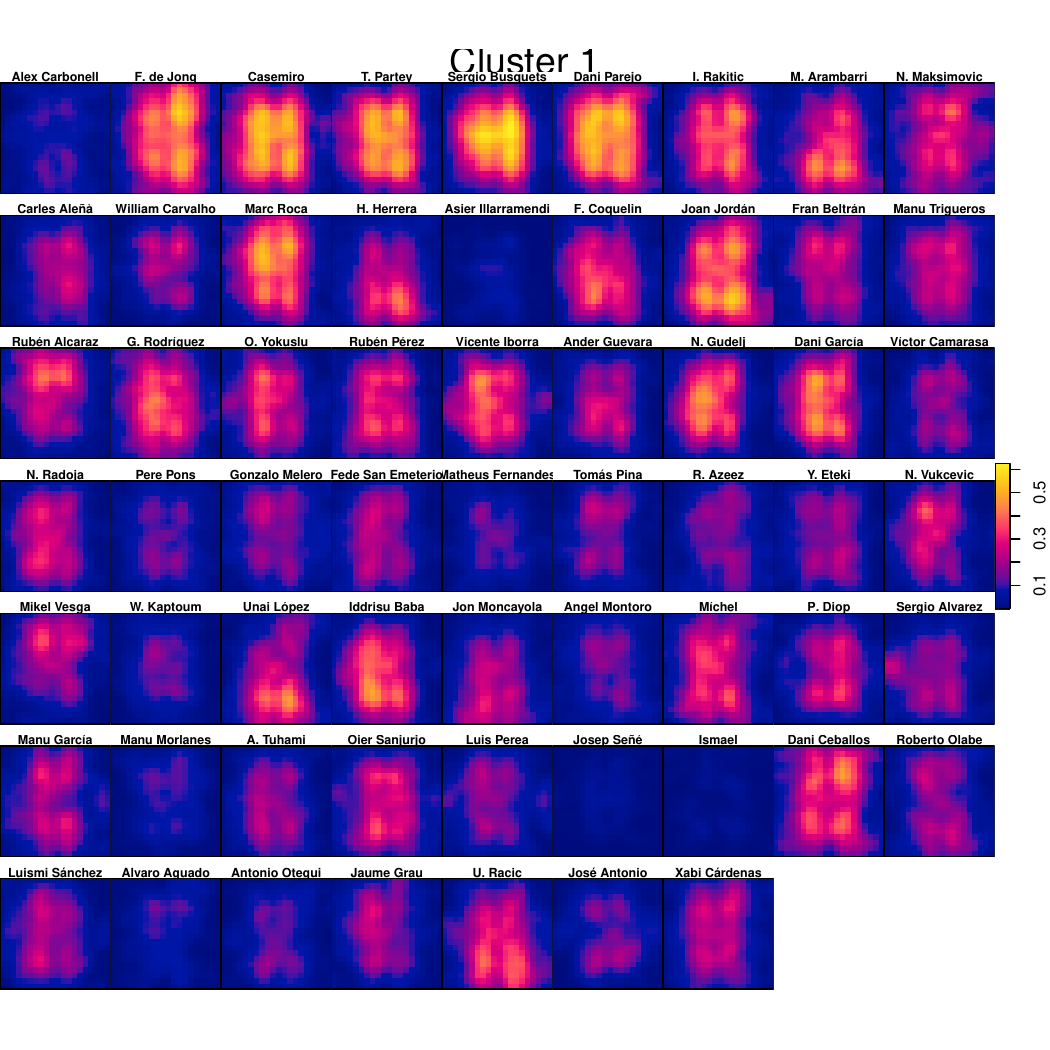}
\includegraphics[scale=0.9]{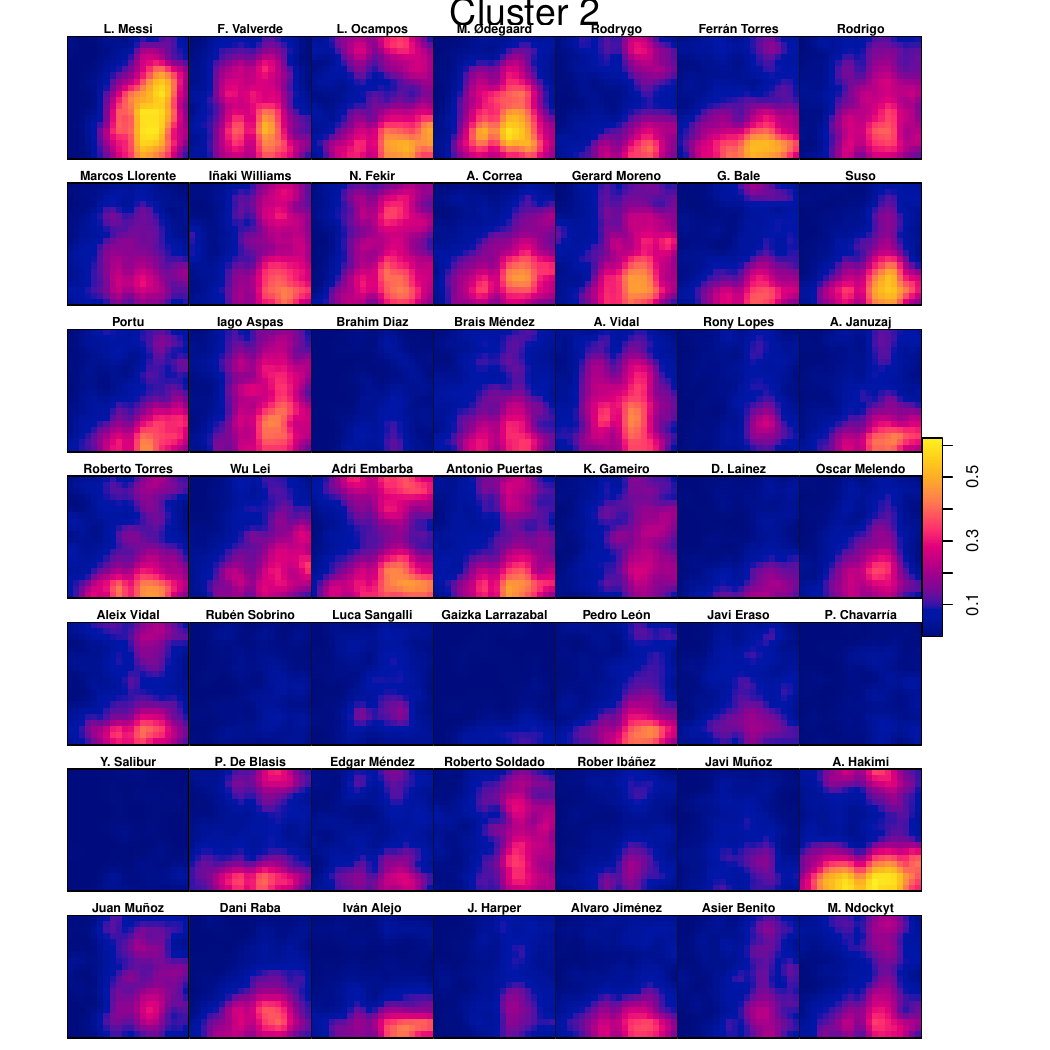}
\includegraphics[scale=0.9]{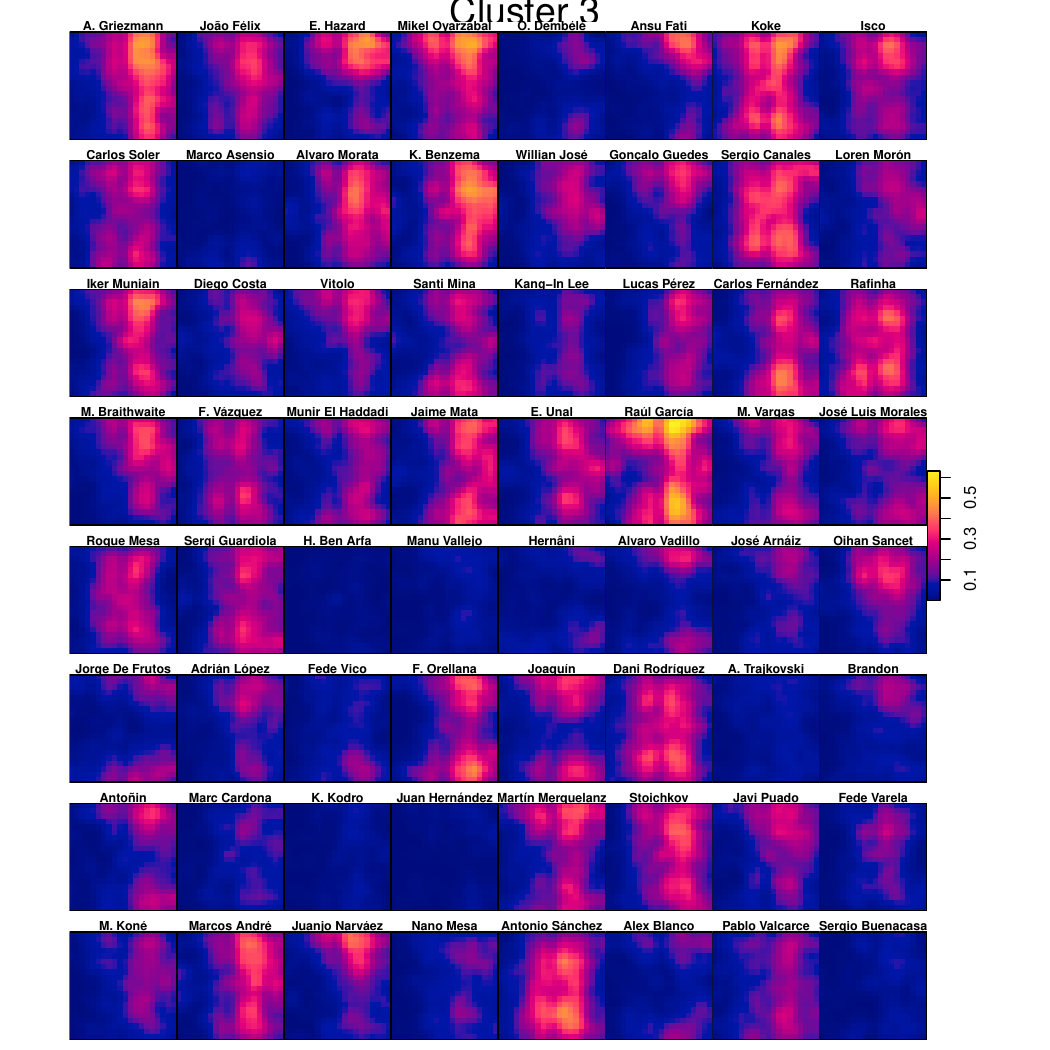}
\includegraphics[scale=0.9]{cluster-04.pdf}
\includegraphics[scale=0.9]{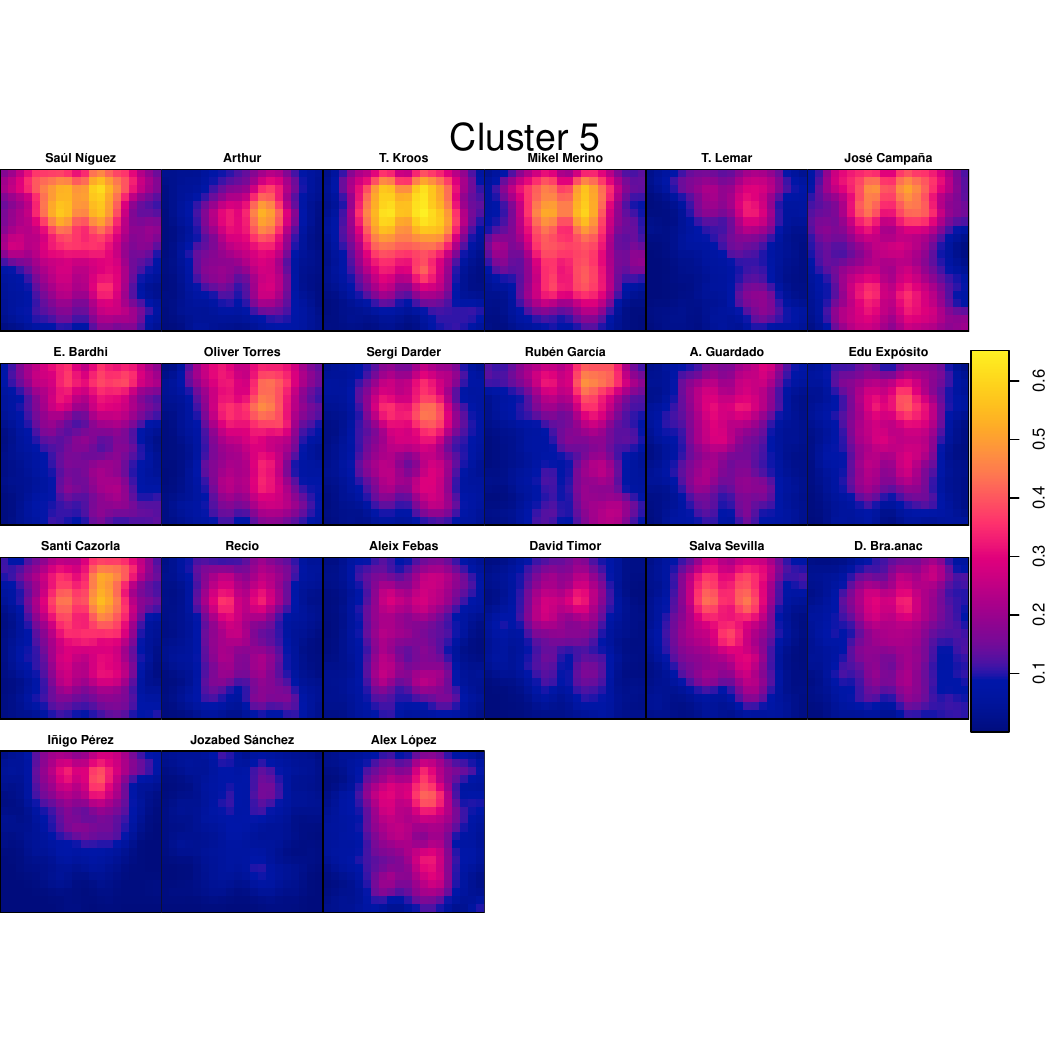}
\includegraphics[scale=0.9]{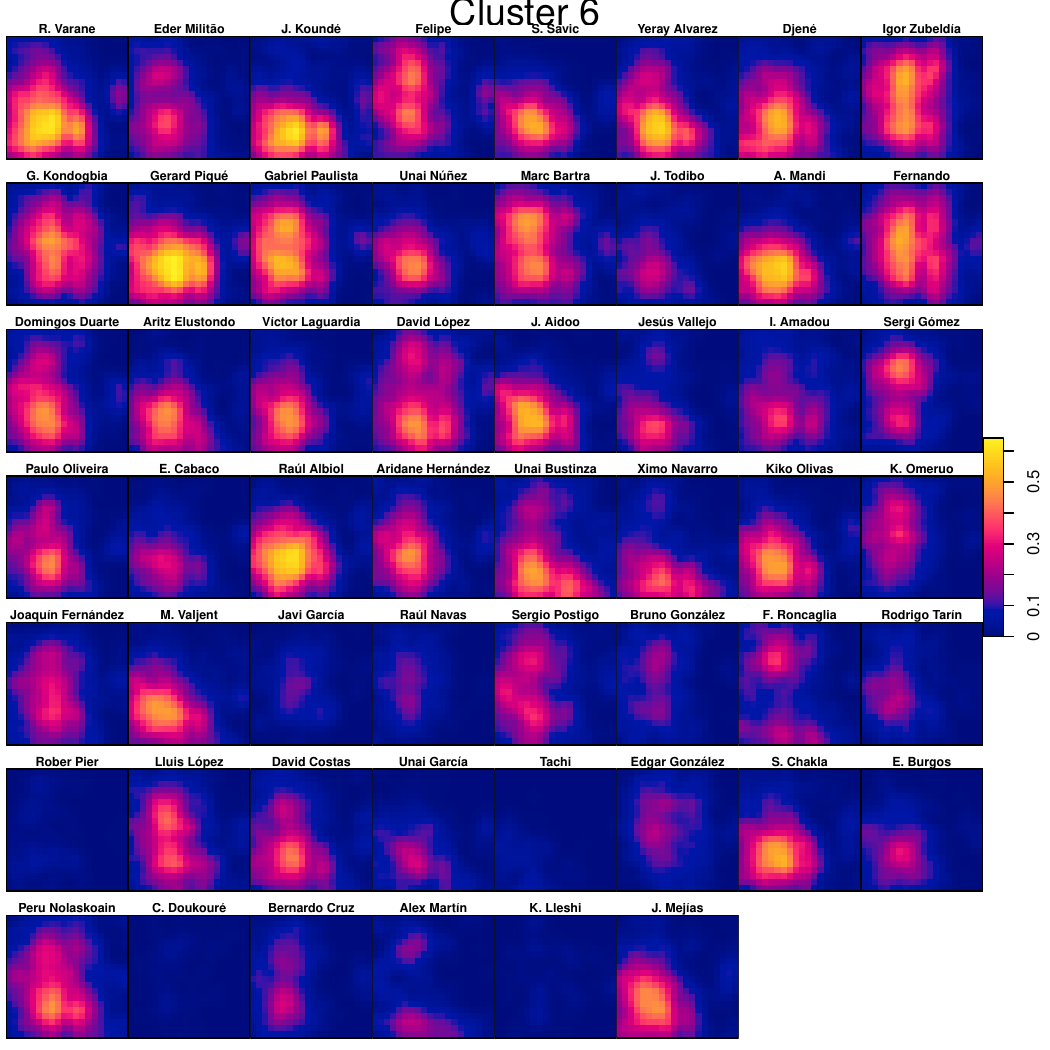}
\includegraphics[scale=0.9]{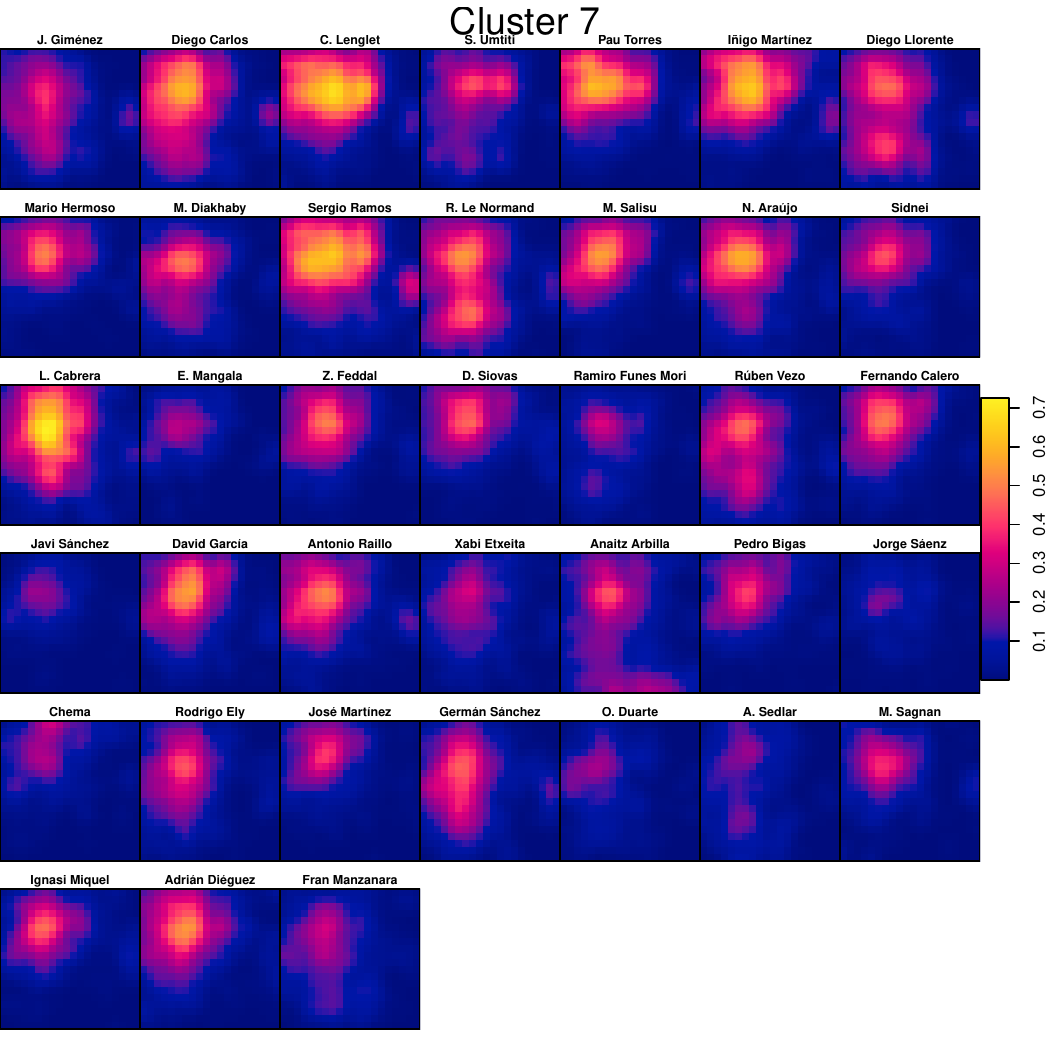}
\includegraphics[scale=0.9]{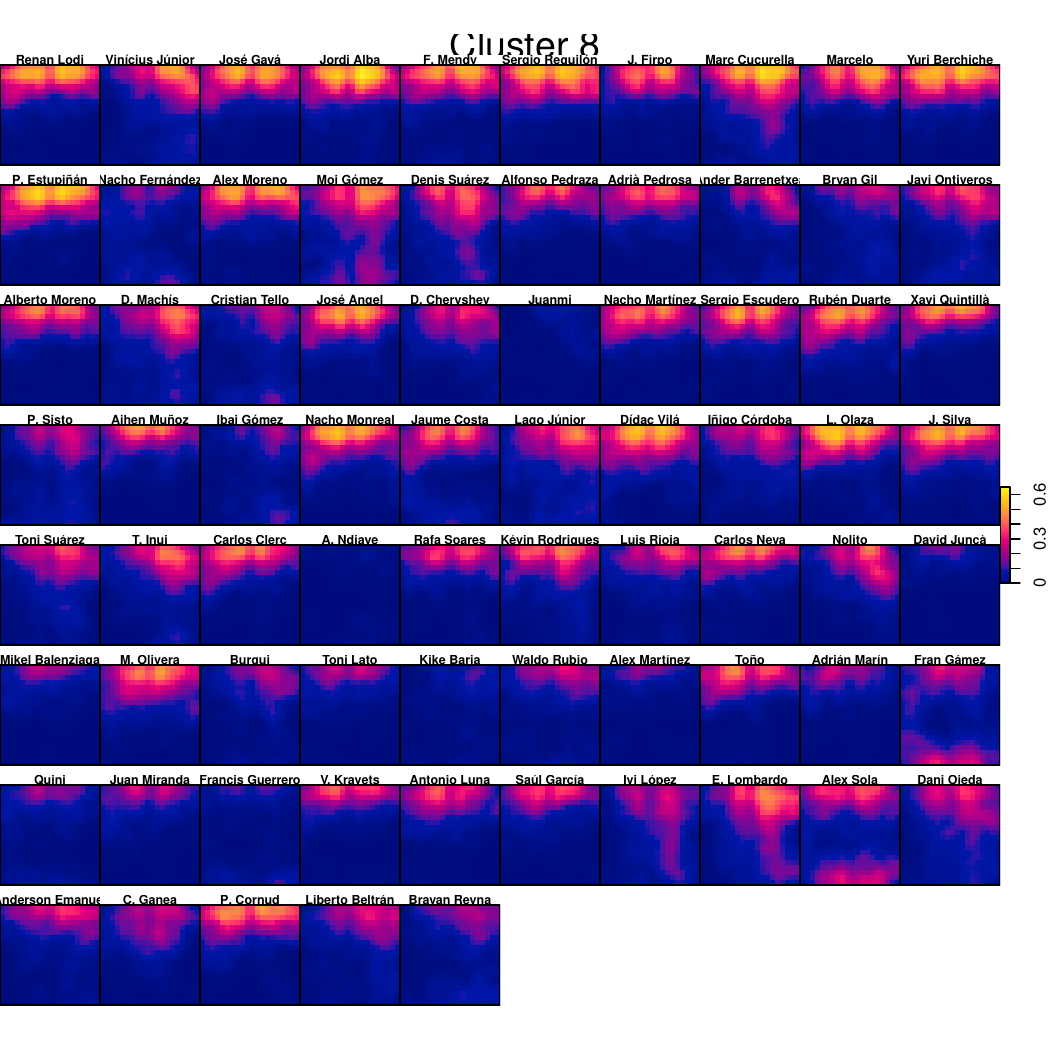}
\includegraphics[scale=0.9]{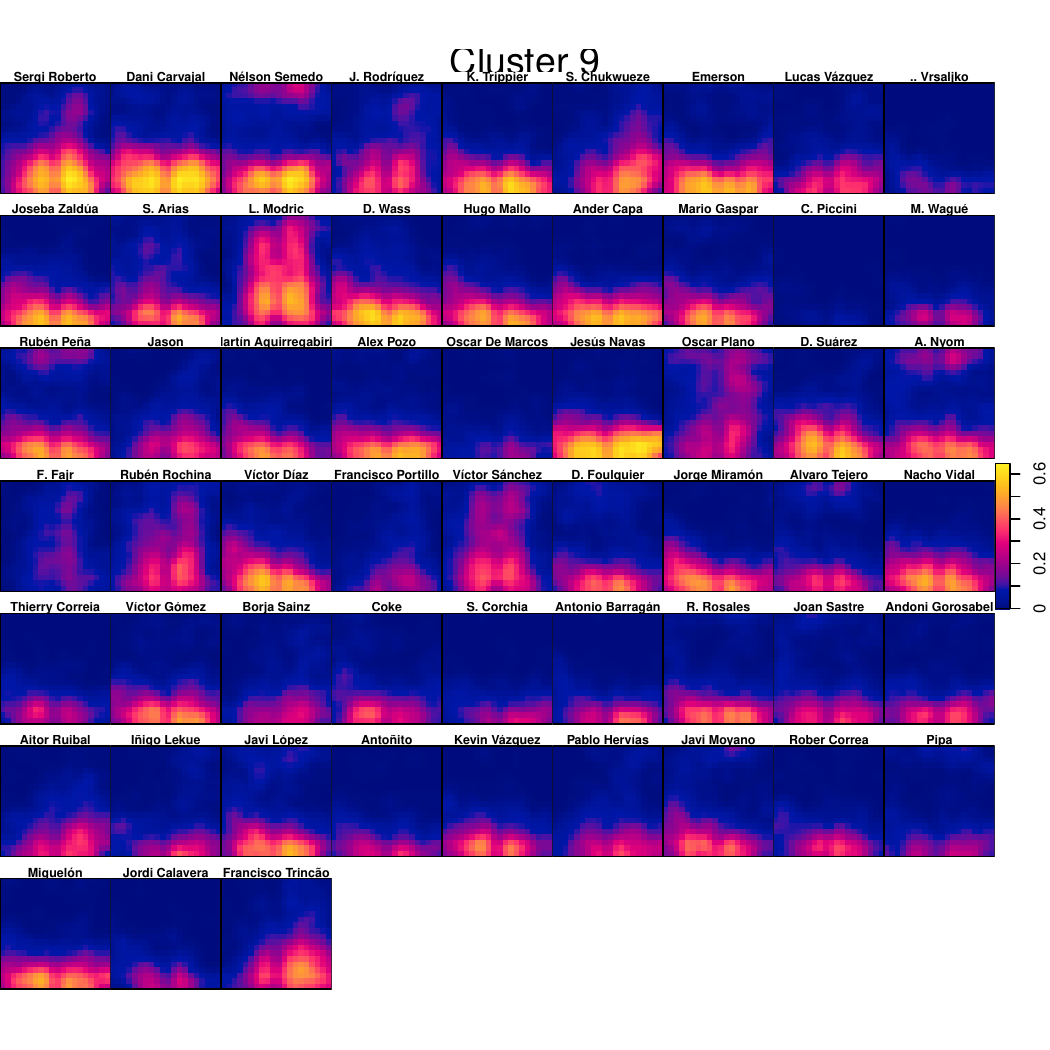}
\includegraphics[scale=0.9]{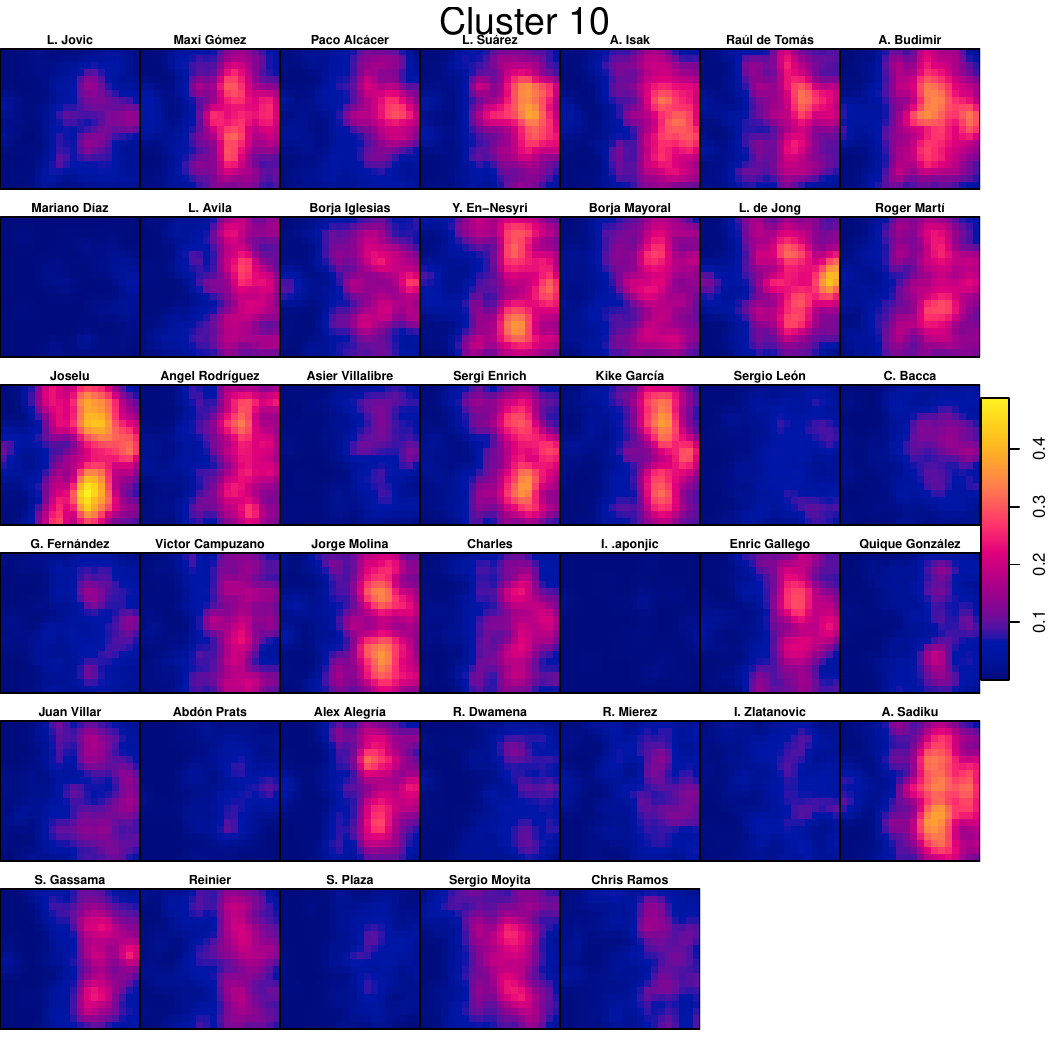}

\end{document}